# A Feasibility Study on Opportunistic Rainfall Measurement From Satellite TV Broadcasts

*F. Sapienza, G. Bacci, F. Giannetti, V. Lottici, A. Vaccaro, G. Serafino, A. Ortolani, F. Caparrini, A. Antonini, S. Melani, A. Mazza, L. Baldini, E. Adirosi, and L. Facheris*

*Abstract* – Rainfall precipitation maps are usually derived based on the measurements collected by classical weather devices, such as rain gauges and weather stations. This article aims to show the benefits obtained by opportunistic rainfall measurements based on signal strength measurements provided by commercial-grade satellite terminals (e.g., used in TV broadcasting). To assess not only the feasibility of this approach, with significant advantages in terms of capital and operational expenditure, but also improvements in terms of accuracy, we focus on a case study for agricultural applications using a Gaussian-modeled synthetic rain over a specific, real-world test area.

## 1. Introduction and Motivation

Monitoring of precipitation in the territory is gaining a paramount importance in many fields of applications, including public safety, business-oriented services (e.g., agriculture), and personal services (e.g., leisure activities) [1]. As an example, accurate precipitation maps allow the agricultural sector to make strategic decisions to improve the quality of production and manage risks [2].

To be effective, estimation of precipitation on the ground must be provided through precipitation maps that must be characterized by 1) accuracy in measuring precipitation rates; 2) completeness, continuity, and high spatial-temporal resolution; 3) negligible delays in map generation compared to measurement acquisition. This has been traditionally accomplished by combining measurements collected by professional instruments, such as weather stations, weather radar, disdrometers, and rain gauges (RGs). In the past few decades, microwave ($\mu$W) links have been investigated to opportunistically retrieve precipitation estimates by correlating radio-frequency attenuation and precipitation along the propagation path.

This includes satellite-based rain-sensing devices, which typically exploit geosynchronous earth orbit (GEO) satellites for TV broadcasting at the Ku-band (10 ÷ 13 GHz). This approach shows the following advantages [3]: 1) no need for authorization, as satellite downlink (DL) signals can be freely acquired and processed; 2) deployment of receiving terminals with minimal technical requirements; and 3) the possibility to use inexpensive commercial-grade devices for domestic reception.

This article aims to show the feasibility of using satellite-based rain-sensing devices to augment the capillarity of a pluviometric network so as to increase the accuracy and the resolution of rainfall maps. For simulation purposes, this study takes advantage of 1) a synthetic rain, modeled as a circularly Gaussian precipitation field, and 2) a classical spatialization method based on the inverse distance weighting (IDW) interpolation [4]. Although the literature offers a number of more elaborate and performing spatialization techniques (e.g., [5]), here we adopt a simpler approach to better emphasize the potential of a satellite-based sensor network.

Fabiola Sapienza, Giacomo Bacci, Filippo Giannetti, and Vincenzo Lottici are with the Dipartimento Ingegneria dell'Informazione, University of Pisa, Via G. Caruso 16, 56122 Pisa, Italy; e-mail: fabiola.sapienza@ing.unipi.it, giacomo.bacci@unipi.it, filippo.giannetti@unipi.it, vincenzo.lottici@unipi.it.

Attilio Vaccaro and Giovanni Serafino are with MBI srl, Via F. Squartini 7, 56121 Pisa, Italy; e-mail: avaccaro@mbigroup.it, gserafino@mbigroup.it.

Alberto Ortolani, Francesca Caparrini, Andrea Antonini, Samantha Melani, and Alessandro Mazza are with Consorzio LaMMA c/o CNR-IBE, Via Madonna del Piano 10, 50019 Sesto Fiorentino, Italy; e-mail: ortolani@lamma.toscana.it, caparrini@lamma.toscana.it, antonini@lamma.toscana.it, melani@lamma.toscana.it, mazza@lamma.toscana.it.

Luca Baldini and Elisa Adirosi are with the Institute of Atmospheric Sciences and Climate, National Research Council of Italy, Via Fosso del Cavaliere 100, 00133 Rome, Italy; e-mail: l.baldini@isac.cnr.it, elisa.adirosi@isac.cnr.it.

Luca Facheris is with the Dipartimento Ingegneria dell'Informazione, University of Florence, Via di S. Marta 3, 50139 Florence, Italy; e-mail: luca.facheris@unifi.it.

## 2. System Model

### 2.1 Basic Concepts of Satellite-Based Rain Sensing

The opportunistic rain-sensing method adopted in this study is based on measurements of the $\mu$W link attenuation and exploits the popular power law [3]

$$A/L = \gamma = \alpha \cdot R^\beta \qquad (1)$$

where $A$ (in dB) represents the total additional attenuation experienced by the signal strength level during the rain (*wet* condition) with respect to the pre-rain value (*dry* condition), $L$ (in km) is the length of the *wet* radio path, $\gamma$ (in dB/km) is the relevant specific attenuation, $R$ (in mm/h) is the rain rate (assumed constant along the path $L$), and $\alpha$ and $\beta$ are frequency- and polarization-dependent empirical coefficients, respectively [3].

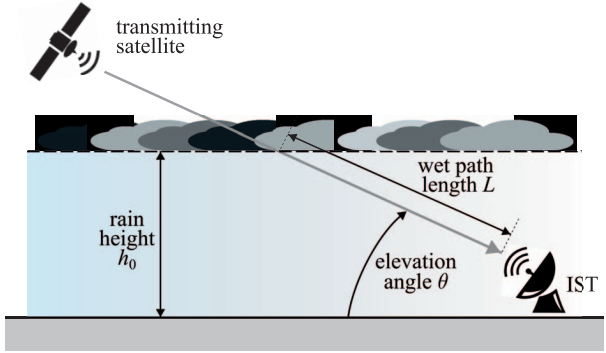

Figure 1. Geometry of a slanted satellite wet link.

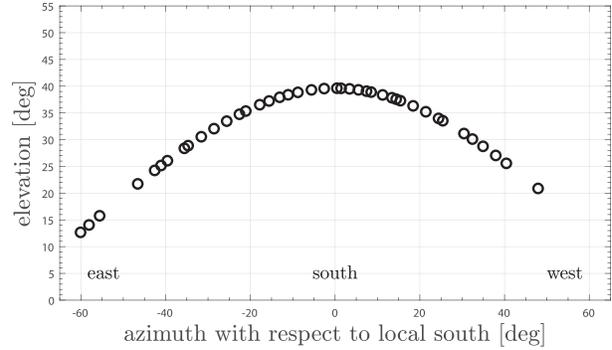

Figure 2. GEO arc as seen from Pisa, Italy.

The length $L$ can be derived considering the link geometry, shown in Figure 1 for a slanted satellite wet link through rain, assuming the tropospheric model that is detailed in [6]. The satellite elevation $\theta$ is a known design parameter depending on the receiver location. The rain height $h_0$ can be derived from the 0°C isotherm height provided either by a short-term forecast based on weather models or based on averages taken at the receive location or by using the methodology proposed in [6].

The attenuation $A$ can be derived through the signal-to-noise ratio (SNR) $\eta = E_s/N_0$ provided by most commercial receivers as the ratio between the energy per symbol $E_s$ and the one-sided noise power spectral density $N_0$. In particular, $A = (\eta^{\text{dry}}/\eta^{\text{wet}})(1 - \xi) + \xi$, where the superscript "dry" denotes the reference level (suitably pre-stored during clear-sky days), the superscript "wet" denotes the current reading during precipitation, and $\xi$ is a design parameter dependent on noise contributions (see [7, eq. (4)]). Note that accurate rain sensing requires suitably adapted relationships derived from (1) to account for the layered structure of the troposphere (e.g., including the melting layer) [7].

### 2.2 Multi-Satellite Rain Sensing

Based on the principles outlined in Section 2.1, we can opportunistically exploit the satellite signals used for direct-to-home (DTH) TV broadcasting. In particular, at any site, there are usually several DTH TV broadcasting satellites, in line-of-sight on the GEO arc (i.e., the Clarke belt as seen in the sky from Earth) with a footprint covering the receive location. For example, assuming Pisa, Italy (10.4147°E, 43.7117°N), as the receive location, there are about 40 broadcasting satellites in the Ku band, lying on the GEO arc from 70.5°E to 37.5°W, with at least one beam covering the city [3] (Figure 2).

A practical implementation of a multi-satellite rain sensor (MSRS) consists of resorting to a commercial-grade multi-beam antenna for DTH, such as the one depicted in Figure 3, which can allocate up to 16 low-noise block converters (LNBs) spanning over an arc range of 40°. A more detailed description of the MSRS can be found in [8]. Each LNB is able to measure the SNR $\eta$, used to estimate the rainfall rates based on the model described in Section 2.1, which contribute to determine the rainfall maps, using the methods outlined in Section 3.

### 2.3 Test Area

To conduct the feasibility study, we focus on a specific case study, given by a 12 km² × 12 km² region near the city of Pisa, Tuscany, Italy, as shown in Figure 4 (red box). This area hosts five RG (three belonging to the Tuscan regional hydro-pluviometric network and accessed through the LaMMA regional network [10] and two from the ETG network [11]) and two MSRSs with coordinates reported in Table 1. Each MSRS accommodates two LNBs. In particular, the LNBs of *Podere Rottaia* point toward the satellites *Eutelsat* 10A at 10°E and *MonacoSat* at 52°E, respectively, whereas the LNBs of *Centro Avanzi* point toward the satellites

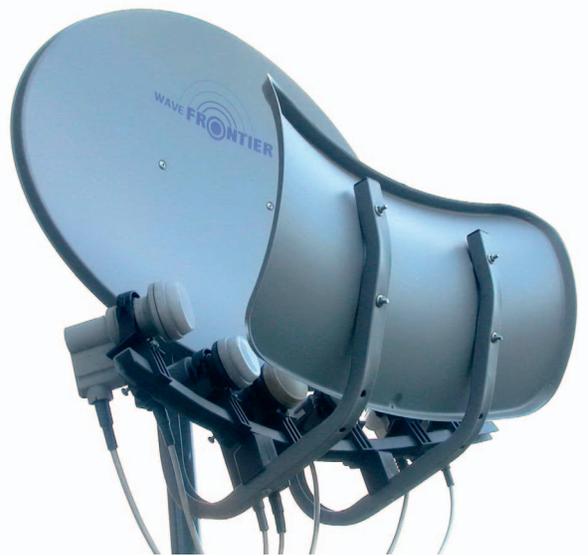

Figure 3. A close-up of the multiple-LNB receiver, using the *Wave-Frontier T90* dual reflector toroidal antenna [9].

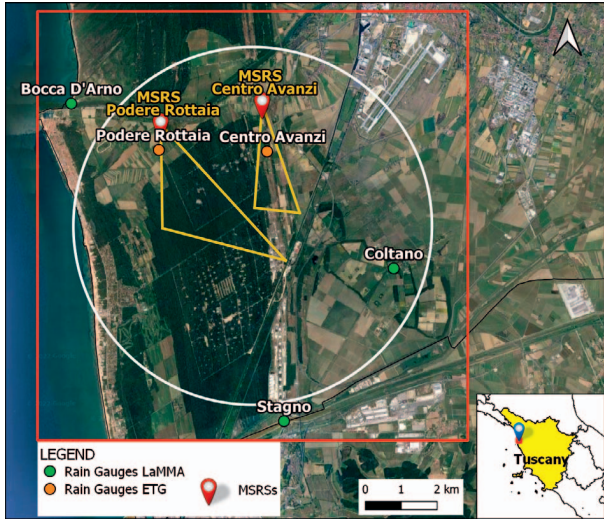

Figure 4. Locations of rain sensors in the test area.

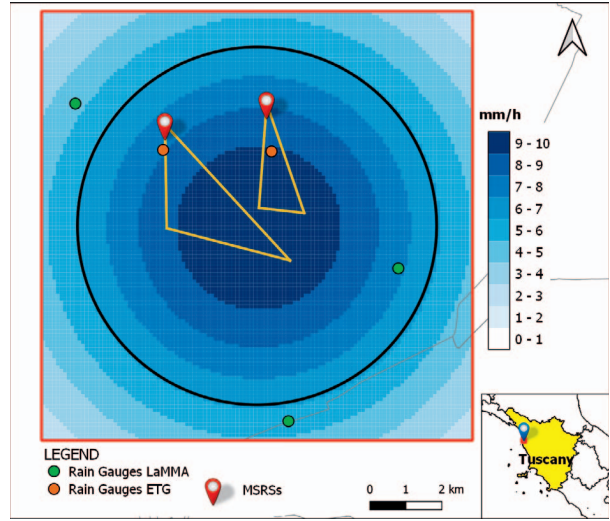

Figure 5. Simulated rain model.

*Astra 4A* at 4.8°E and *Astra 2G* at 28.2°E, respectively. The ground projections of the slanted paths are indicated with yellow lines, assuming a rain height of 2,700 m. Since the reliability of data collected by MSRSs is still under investigation, we propose to use the simulated rain model detailed in Section 3, with all the rain sensors reported in Figure 4 considered capable of retrieving the synthetic rain model.

## 3. Interpolation Method

The simulated rain model is the circular Gaussian precipitation field depicted in Figure 5, whose rain rate $R$, as a function of the coordinates $(x, y)$, follows the relation

$$R(x,y) = \bar{R} \cdot \exp\left\{-\frac{(x-x_0)^2 + (y-y_0)^2}{2\sigma^2}\right\} \quad (2)$$

where the origin of the system coordinates $(x=0, y=0)$ is assumed to be located in the northwest vertex (10.2691°E, 43.7040°N), while $x_0 = -y_0 = 6$ km, $\bar{R} = 10$ mm/h, and $\sigma = 5$ km. This corresponds to a significant rain event focused on the central sector of the testing area, presented as a circular area in Figures 4 and 5. Note that no specific duration is associated to the simulated event, and the analysis considered in this article is valid regardless of the event duration since it is based on the instantaneous rain rate.

A spatialization technique is applied to the synthetic rain rate measured by rain-sensing devices in order to retrieve the precipitation field. The study area is divided into $120 \times 120$ boxes, each with size 100 m $\times$ 100 m. The interpolation method used to estimate the rain rate $\hat{R}_k$ in the $k$th box is the IDW interpolation [4], which computes the rain rate in the grid boxes as a weighted average of the values available at the known points, according to

$$\hat{R}_k = \left(\sum_{n=1}^{N}\frac{R(x_n, y_n)}{d_{n,k}^4}\right) \bigg/ \left(\sum_{n=1}^{N}\frac{1}{d_{n,k}^4}\right) \quad (3)$$

where $N$ is the number of the rain sensors, $R(x_n, y_n)$ is the rain rate measured using (2) by the $n$th device located at coordinates $(x_n, y_n)$, and $d_{n,k}$ is the distance between the $k$th box and $(x_n, y_n)$. Note that values closer to known points have a larger weight compared to farther values.

## 4. Numerical Results

To retrieve the precipitation field in the area of interest, we focus on two scenarios: 1) synthetic measurements from RGs only and 2) synthetic measurements from the three RGs from the LaMMA network and two MSRSs, placed in the same locations as the two RGs, from the ETG network. The obtained rainfall maps are reported in Figures 6 and 7, respectively. With reference to Figure 7, each LNB of the two MSRSs is considered as a virtual RG (red dots) located at the end on the ground projections of the slanted path for the satellites detailed in Section 2.3 and measuring the average rain rate observed along its slanted path. Since we are interested in the circularly modeled rain event simulated according to (2) and reported in Figure 5, we focus on the central sector, delimited by the black circle in Figures 6 and 7. As can be seen, the usage of the two MSRSs improves the rainfall map, as the rain field appears to better reflect the synthetic event in the area of interest.

Table 1. WGS84 coordinates of the rain-sensing devices

| (Device type) ID | Long. (°E) | Lat. (°N) |
|---|---|---|
| (RG) Bocca D'Arno | 10.2803 | 43.6807 |
| (RG) Podere Rottaia | 10.3104 | 43.6687 |
| (RG) Centro Avanzi | 10.3104 | 43.6687 |
| (RG) Coltano | 10.3909 | 43.6379 |
| (RG) Stagno | 10.3104 | 43.5999 |
| (MSRS) Podere Rottaia | 10.3112 | 43.6751 |
| (MSRS) Centro Avanzi | 10.3464 | 43.6800 |

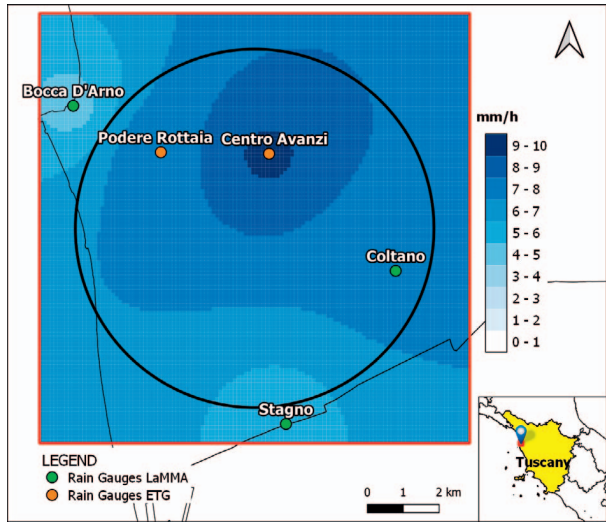

Figure 6. Rainfall map on the test area retrieved by using only RGs (scenario A).

Table 2. Rain rate RMSE (mm/h)

|  | Scenario A (RG only) | Scenario B (RG + MSRS) |
| --- | --- | --- |
| Extended area | 1.7662 | 1.9294 |
| Central sector | 1.1294 | 0.7997 |

To quantitatively compare the two scenarios, we compute the root mean square error (RMSE) for the two rainfall maps, considering both the black-contoured circular area and the extended, red-contoured squared region. Table 2 shows the numerical results, which can be interpreted as follows. On one side, using the proposed MSRS improves the RMSE of the rainfall map in the area of interest (black-contoured circular area): this paves the way to include this kind of device in the set of rainfall estimation sensors. On the other side, the increased RMSE measured in the red-contoured squared region, which is due to larger estimation errors close to the area vertices (and hence outside the region of interest), gives us a design criterion for the MSRS deployment: the right path to pursue is to densify the areas of interests with this kind of sensor, which also shows all the advantages (e.g., low device cost, usage of opportunistic signals, and agile installation) summarized in Section 1.

## 5. Conclusions

The test carried out demonstrated the potential of multi-satellite sensor systems as rain sensors for rainfall map retrieval to reduce deployment costs while increasing estimation accuracy. Note also that the usage of multiple satellites increases the resilience of the network against occasional issues (e.g., signal interruption and satellite re-positioning) on a specific satellite system. The proposed MSRS-based system is under deployment in agricultural environments in Tuscany for experimental tests within the INSIDERAIN project [12]. This also represents a preliminary phase of a broad campaign of experimental field trials to be carried out in the framework of the EU-funded SCORE project [13], which aims to increase climate resilience in European coastal cities.

## 6. Acknowledgments

This work was supported by the project INSIDERAIN funded by Tuscany regional administration, Italy, Decreto no. 21885, December 18, 2020, and by the project SCORE funded by European Commission's Horizon 2020 research and innovation program under grant agreement no. 101003534. This article is also based on work from COST Action CA20136 OPEN-SENSE, supported by COST (European Cooperation in Science and Technology).

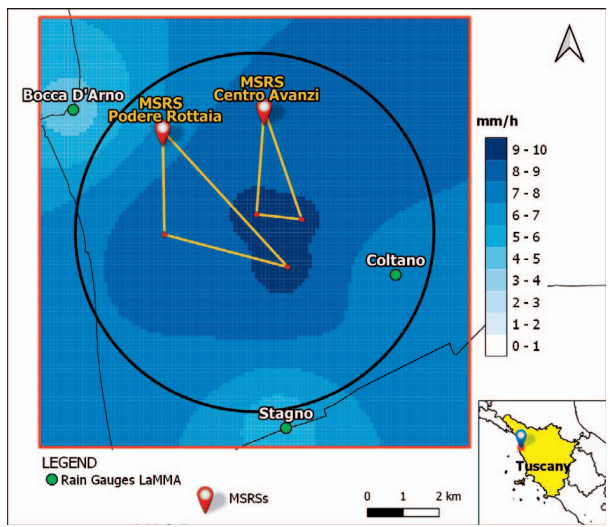

Figure 7. Rainfall map on the test area retrieved by combining RGs and MSRSs (scenario B).